\documentclass[letterpaper]{mc2023}
\usepackage{tabls}
\usepackage{cites}
\usepackage{epsf}
\usepackage{appendix}
\usepackage{ragged2e}
\usepackage[top=1in, bottom=1in, left=1in, right=1in]{geometry}
\usepackage{enumitem}
\setlist[itemize]{leftmargin=*}
\usepackage{caption}
\title{Sensitivity Analysis and Uncertainty Quantification on \\ Point Defect Kinetics Equations with Perturbation Analysis}
\author{%
  \textbf{Miaomiao ~Jin$^1$, and Jilang ~Miao$^1$}\footnote{Corresponding author (jlmiao@psu.edu)}\vspace{3pt} \\
  $^1$ 
 Department of Nuclear Engineering, The Pennsylvania State University, \\ University Park, 16802, PA, USA \vspace{6pt}\\ 
}
\newcommand{\authorHead}{Miaomiao Jin \& Jilang Miao}
\newcommand{\shortTitle}{Sensitivity Analysis and Uncertainty Quantification on  Point Defect Kinetics Equations with Perturbation Analysis}
\begin{document}
\setlength{\abovedisplayskip}{0pt}
\setlength{\belowdisplayskip}{0pt}
\maketitle


\pagestyle{fancy} \cfoot{\thepage}
\fancypagestyle{firstpage}{\fancyhead[C]{\footnotesize{\emph{
M\&C 2023 - The International Conference on Mathematics and Computational Methods Applied \\
to Nuclear Science and Engineering $\cdot$ Niagara Falls, Ontario, Canada $\cdot$ August 13 -- 17, 2023}}}
\cfoot{}}
\thispagestyle{firstpage}
\fancyhead[CE]{{\scriptsize \authorHead}}
\fancyhead[CO]{{\scriptsize \shortTitle}}

\justify 
\begin{abstract}
The concentration of radiation-induced point defects in general materials under irradiation is commonly described by the point defect kinetics equations based on rate theory. However, the parametric uncertainty in describing the rate constants of competing physical processes such as recombination and loss to sinks can lead to a large uncertainty in predicting the time-evolving point defect concentrations. Here, based on the perturbation theory, we derived up to the third order correction to the solution of point defect kinetics equations. This new set of equations enable a full description of continuously changing rate constants, and can accurately predict the solution up to $50\%$ deviation in these rate constants. These analyses can also be applied to reveal the sensitivity of solution to input parameters and aggregated uncertainty from multiple rate constants. 
\end{abstract}
\section{INTRODUCTION} 
Radiation-induced defects are the key to the degradation of materials properties such as segregation, swelling and embrittlement \cite{was2016fundamentals}. Comparing to the thermal equilibrium condition, a much higher concentration of crystalline defects can be created due to high-energy radiation particles colliding with lattice atoms. As defects can significantly accelerate the rates of diffusion and reaction, a description of defect concentration in materials under the radiation environment constitutes the basis to predictive modeling of radiation effects. In particular, point defects (vacancy and interstitial) are commonly described by the point defect kinetics equations via the chemical rate theory \cite{was2016fundamentals}. Under the concept of mean field rate theory where the spatial dependence is neglected, the change in defect concentration can be described from several competing processes, including direct defect production from irradiation, vacancy-interstitial recombination, and defect loss to sinks such as dislocations and grain boundaries. Mathematically,
\begin{equation}
    \frac{\mathrm{d}}{\mathrm{d} t}\left(\begin{array}{c}C_{\mathrm{v}} \\ C_{\mathrm{i}}\end{array}\right) = K_0 -K_{\mathrm{iv}} C_{\mathrm{i}}  C_{\mathrm{v}}   -C_{\mathrm{s}}\left(\begin{array}{c}K_{\mathrm{vs}} C_{\mathrm{v}} \\ K_{\mathrm{is}} C_{\mathrm{i}}\end{array}\right) 
    \label{eq::pk}
\end{equation}
where $C_{\mathrm{v}}$ and $C_{\mathrm{i}}$ are vacancy and interstitial concentration, respectively, $K_0$ is defect production rate, $K_{\mathrm{iv}}$ is vacancy–interstitial recombination rate constant, $K_{\mathrm{vs}}$ and $K_{\mathrm{is}}$ are vacancy–sink and interstitial-sink reaction rate constant, respectively. The values of these rate constants are hence of significance to solving the concentrations. Physically, such rates can be derived via diffusion- or reaction-limited analysis, which yields a formulation depending on the defect interactions and mobilities \cite{was2016fundamentals}. As a typical methodology, lower length scale computational methods (e.g., density functional theory (DFT) and molecular dynamics) are used to determine fundamental quantities such as interaction strength and diffusion energy barriers. Such treatment inevitably introduces uncertainly in the rate parameters due to several reasons: i) lower length scale methods have its own accuracy limit due to calculation settings and potential choices; ii) limited defect migration pathways are considered due to complexity; and iii) pre-existing damage is hardly captured to modify the current defect energetics. Short et al. demonstrated that a slight change in the vacancy migration energy barrier (0.03 eV) can cause drastic changes in the point defect concentration profile in self-ion irradiation alpha-Fe \cite{short2016modeling}. It is thus of significance to perform parametric sensitivity analysis and uncertainty quantification due to the avoidable uncertainties in input parameters. Although one may modify the parameter in little increment/decrement to solve point defect kinetics, it is generally time-consuming and can not exhibit a full picture of uncertainty variation in the nearby parameter regions with respect to the parameters in use. To tackle this problem, we use perturbation theory~\cite{nayfeh2008perturbation} to derive a new set of equations, which can be solved in concurrent with Eq. \ref{eq::pk}, and the uncertainty of defect concentration can be nicely captured by the correction terms. These analyses can be combined to yield a multi-parameter uncertainty quantification considering the joint distribution of the rate parameters.     
\section{PERTURBATION ANALYSIS} 
\label{sec:pert}
Consider a perturbation to an input parameter $K$ (e.g. $K_0$, $K_{\mathrm{iv}}$) in Eq~\ref{eq::pk} in the form of 
\begin{equation}
    K \rightarrow K(1+\epsilon)
    \label{eq::K(1+e)}
\end{equation}
then the solution can be expressed in the form of perturbative expansion, 
\begin{equation}
\left(\begin{array}{c}C_{\mathrm{v}}  \\ C_{\mathrm{i}} \end{array}\right) = 
\left(\begin{array}{c}C_{\mathrm{v}}^{(0)} \\ C_{\mathrm{i}}^{(0)}\end{array}\right) + 
\epsilon 
\left(\begin{array}{c}C_{\mathrm{v}}^{(1)} \\ C_{\mathrm{i}}^{(1)}\end{array}\right) +    
\epsilon^2
\left(\begin{array}{c}C_{\mathrm{v}}^{(2)} \\ C_{\mathrm{i}}^{(2)}\end{array}\right) +    
\epsilon^3
\left(\begin{array}{c}C_{\mathrm{v}}^{(3)} \\ C_{\mathrm{i}}^{(3)}\end{array}\right) +   
\cdot \cdot \cdot
\label{eq::y+y1+...}
\end{equation}
where $C_\mathrm{v}^{(0)}$ and $C_\mathrm{i}^{(0)}$ are the solution of the unperturbed Eq~\ref{eq::pk}. The correction terms ($C_\mathrm{v}^{(1)}$, $C_\mathrm{i}^{(1)}$, $C_\mathrm{v}^{(2)}$, $C_\mathrm{i}^{(2)}$, etc.) can be found by
substituting Eq~\ref{eq::K(1+e)} and Eq~\ref{eq::y+y1+...} into Eq~\ref{eq::pk} and matching the coefficients of the perturbation $\epsilon$. 
The equations needed to find results up to the thrid order are listed below. 
\subsection{Differential equations of higher order solution corrections}
\label{sec::c123}
\subsubsection{Perturbation on $K_{0}$ }
\label{sec::K0}
The equations to solve for the change due to $K_{0}$ are given in Eq~\ref{eq::K01} to Eq~\ref{eq::K03}. 
\begin{equation}
    \frac{\mathrm{d}}{\mathrm{d} t}\left(\begin{array}{c}C_{\mathrm{v}}^{(1)} \\ C_{\mathrm{i}}^{(1)}\end{array}\right) = K_0 -K_{\mathrm{iv}} \left ( C_{\mathrm{i}}^{(0)} C_{\mathrm{v}}^{(1)} +  C_{\mathrm{v}}^{(0)} C_{\mathrm{i}}^{(1)} \right ) -C_{\mathrm{s}}\left(\begin{array}{c}K_{\mathrm{vs}} C_{\mathrm{v}}^{(1)} \\ K_{\mathrm{is}} C_{\mathrm{i}}^{(1)}\end{array}\right) 
    \label{eq::K01}
\end{equation}
\begin{align}
\begin{split}
    \frac{\mathrm{d}}{\mathrm{d} t}\left(\begin{array}{c}C_{\mathrm{v}}^{(2)} \\ C_{\mathrm{i}}^{(2)}\end{array}\right) 
    & =-K_{\mathrm{iv}}\left(C_{\mathrm{i}}^{(0)} C_{\mathrm{v}}^{(2)}+C_{\mathrm{v}}^{(0)} C_{\mathrm{i}}^{(2)}+C_{\mathrm{i}}^{(1)} C_{\mathrm{v}}^{(1)}\right) 
     -C_{\mathrm{s}}\left(\begin{array}{c}K_{\mathrm{vs}} C_{\mathrm{v}}^{(2)} \\ K_{\mathrm{is}} C_{\mathrm{i}}^{(2)}\end{array}\right) 
\end{split}
\label{eq::K02}
\end{align}
\begin{align}
\begin{split}
\frac{\mathrm{d}}{\mathrm{d} t}\left(\begin{array}{c}C_{\mathrm{v}}^{(3)} \\ C_{\mathrm{i}}^{(3)}\end{array}\right)
&=-K_{\mathrm{iv}}\left(C_{\mathrm{i}}^{(0)} C_{\mathrm{v}}^{(3)}+C_{\mathrm{v}}^{(0)} C_{\mathrm{i}}^{(3)}+C_{\mathrm{i}}^{(1)} C_{\mathrm{v}}^{(2)}+C_{\mathrm{v}}^{(1)} C_{\mathrm{i}}^{(2)}\right) 
 -C_{\mathrm{s}}\left(\begin{array}{c}K_{\mathrm{vs}} C_{\mathrm{v}}^{(3)} \\ K_{\mathrm{is}} C_{\mathrm{i}}^{(3)}\end{array}\right) 
\end{split}
\label{eq::K03}
\end{align}

\subsubsection{Perturbation on $K_{\mathrm{v s}}$ }
\label{sec::Kvs}
The equations to solve for the change due to $K_\mathrm{vs}$ are given in Eq~\ref{eq::Kvs1} to Eq~\ref{eq::Kvs3}. 

\begin{equation}
    \frac{\mathrm{d}}{\mathrm{d} t}\left(\begin{array}{c}C_{\mathrm{v}}^{(1)} \\ C_{\mathrm{i}}^{(1)}\end{array}\right) = -K_{\mathrm{iv}} \left ( C_{\mathrm{i}}^{(0)} C_{\mathrm{v}}^{(1)} +  C_{\mathrm{v}}^{(0)} C_{\mathrm{i}}^{(1)} \right ) -C_{\mathrm{s}}\left(\begin{array}{c}K_{\mathrm{vs}} C_{\mathrm{v}}^{(1)} \\ K_{\mathrm{is}} C_{\mathrm{i}}^{(1)}\end{array}\right)-C_{\mathrm{s}}\left(\begin{array}{c}K_{\mathrm{vs}} C_{\mathrm{v}}^{(0)} \\ 0\end{array}\right)
    \label{eq::Kvs1}
\end{equation}
\begin{align}
\begin{split}
    \frac{\mathrm{d}}{\mathrm{d} t}\left(\begin{array}{c}C_{\mathrm{v}}^{(2)} \\ C_{\mathrm{i}}^{(2)}\end{array}\right) 
    & =-K_{\mathrm{iv}}\left(C_{\mathrm{i}}^{(0)} C_{\mathrm{v}}^{(2)}+C_{\mathrm{v}}^{(0)} C_{\mathrm{i}}^{(2)}+C_{\mathrm{i}}^{(1)} C_{\mathrm{v}}^{(1)}\right)  \\
    & -C_{\mathrm{s}}\left(\begin{array}{c}K_{\mathrm{vs}} C_{\mathrm{v}}^{(2)} \\ K_{\mathrm{is}} C_{\mathrm{i}}^{(2)}\end{array}\right)-C_{\mathrm{s}}\left(\begin{array}{c}K_{\mathrm{vs}} C_{\mathrm{v}}^{(1)} \\ 0\end{array}\right)
\end{split}
    \label{eq::Kvs2}
\end{align}
\begin{align}
\begin{split}
\frac{\mathrm{d}}{\mathrm{d} t}\left(\begin{array}{c}C_{\mathrm{v}}^{(3)} \\ C_{\mathrm{i}}^{(3)}\end{array}\right)
&=-K_{\mathrm{iv}}\left(C_{\mathrm{i}}^{(0)} C_{\mathrm{v}}^{(3)}+C_{\mathrm{v}}^{(0)} C_{\mathrm{i}}^{(3)}+C_{\mathrm{i}}^{(1)} C_{\mathrm{v}}^{(2)}+C_{\mathrm{v}}^{(1)} C_{\mathrm{i}}^{(2)}\right) \\
& -C_{\mathrm{s}}\left(\begin{array}{c}K_{\mathrm{vs}} C_{\mathrm{v}}^{(3)} \\ K_{\mathrm{is}} C_{\mathrm{i}}^{(3)}\end{array}\right)-C_{\mathrm{s}}\left(\begin{array}{c}K_{\mathrm{vs}} C_{\mathrm{v}}^{(2)} \\ 0\end{array}\right)
\end{split}
    \label{eq::Kvs3}
\end{align}

\subsubsection{Perturbation on $K_{\mathrm{i s}}$}
\label{sec::Kis}
The equations to solve for the change due to $K_{is}$ are given in Eq~\ref{eq::Kis1} to Eq~\ref{eq::Kis3}. 
\begin{equation}
    \frac{\mathrm{d}}{\mathrm{d} t}\left(\begin{array}{c}C_{\mathrm{v}}^{(1)} \\ C_{\mathrm{i}}^{(1)}\end{array}\right) = -K_{\mathrm{iv}} \left ( C_{\mathrm{i}}^{(0)} C_{\mathrm{v}}^{(1)} +  C_{\mathrm{v}}^{(0)} C_{\mathrm{i}}^{(1)} \right ) -C_{\mathrm{s}}\left(\begin{array}{c}K_{\mathrm{vs}} C_{\mathrm{v}}^{(1)} \\ K_{\mathrm{is}} C_{\mathrm{i}}^{(1)}\end{array}\right)-C_{\mathrm{s}}\left(\begin{array}{c} 0 \\  K_{\mathrm{is}} C_{\mathrm{i}}^{(0)}  \end{array}\right)
\label{eq::Kis1}
\end{equation}

\begin{align}
\begin{split}
    \frac{\mathrm{d}}{\mathrm{d} t}\left(\begin{array}{c}C_{\mathrm{v}}^{(2)} \\ C_{\mathrm{i}}^{(2)}\end{array}\right) 
    & =-K_{\mathrm{iv}}\left(C_{\mathrm{i}}^{(0)} C_{\mathrm{v}}^{(2)}+C_{\mathrm{v}}^{(0)} C_{\mathrm{i}}^{(2)}+C_{\mathrm{i}}^{(1)} C_{\mathrm{v}}^{(1)}\right) \\
  &   -C_{\mathrm{s}}\left(\begin{array}{c}K_{\mathrm{vs}} C_{\mathrm{v}}^{(2)} \\ K_{\mathrm{is}} C_{\mathrm{i}}^{(2)}\end{array}\right)-C_{\mathrm{s}}\left(\begin{array}{c}0 \\  K_{\mathrm{is}} C_{\mathrm{i}}^{(1)}  \end{array}\right)
\end{split}
\label{eq::Kis2}
\end{align}

\begin{align}
\begin{split}
\frac{\mathrm{d}}{\mathrm{d} t}\left(\begin{array}{c}C_{\mathrm{v}}^{(3)} \\ C_{\mathrm{i}}^{(3)}\end{array}\right)
&=-K_{\mathrm{iv}}\left(C_{\mathrm{i}}^{(0)} C_{\mathrm{v}}^{(3)}+C_{\mathrm{v}}^{(0)} C_{\mathrm{i}}^{(3)}+C_{\mathrm{i}}^{(1)} C_{\mathrm{v}}^{(2)}+C_{\mathrm{v}}^{(1)} C_{\mathrm{i}}^{(2)}\right) \\
& -C_{\mathrm{s}}\left(\begin{array}{c}K_{\mathrm{vs}} C_{\mathrm{v}}^{(3)} \\ K_{\mathrm{is}} C_{\mathrm{i}}^{(3)}\end{array}\right)-C_{\mathrm{s}}\left(\begin{array}{c} 0 \\  K_{\mathrm{is}} C_{\mathrm{i}}^{(2)} \end{array}\right)
\end{split}
\label{eq::Kis3}
\end{align}
The corresponding equations for $K_\mathrm{is}$ are identical to those of $k_\mathrm{vs}$ (Eqs~\ref{eq::Kvs1},~\ref{eq::Kvs2} and \ref{eq::Kvs3}) except the last term in each equation, where $K_\mathrm{vs}$ changes are replaced with $K_\mathrm{is}$ changes accordingly. 

\subsubsection{Perturbation on $C_{\mathrm{s}}$}
\label{sec::Cs}
The equations to solve for the change due to $C_\mathrm{s}$ are given in Eq~\ref{eq::Cs1} to Eq~\ref{eq::Cs3}. 

\begin{equation}
    \frac{\mathrm{d}}{\mathrm{d} t}\left(\begin{array}{c}C_{\mathrm{v}}^{(1)} \\ C_{\mathrm{i}}^{(1)}\end{array}\right) = -K_{\mathrm{iv}} \left ( C_{\mathrm{i}}^{(0)} C_{\mathrm{v}}^{(1)} +  C_{\mathrm{v}}^{(0)} C_{\mathrm{i}}^{(1)} \right ) -C_{\mathrm{s}}\left(\begin{array}{c}K_{\mathrm{vs}} C_{\mathrm{v}}^{(1)} \\ K_{\mathrm{is}} C_{\mathrm{i}}^{(1)}\end{array}\right)-C_{\mathrm{s}}\left(\begin{array}{c} 
    K_{\mathrm{vs}} C_{\mathrm{v}}^{(0)} \\  K_{\mathrm{is}} C_{\mathrm{i}}^{(0)}  \end{array}\right)
\label{eq::Cs1}
\end{equation}

\begin{align}
\begin{split}
    \frac{\mathrm{d}}{\mathrm{d} t}\left(\begin{array}{c}C_{\mathrm{v}}^{(2)} \\ C_{\mathrm{i}}^{(2)}\end{array}\right) 
    & =-K_{\mathrm{iv}}\left(C_{\mathrm{i}}^{(0)} C_{\mathrm{v}}^{(2)}+C_{\mathrm{v}}^{(0)} C_{\mathrm{i}}^{(2)}+C_{\mathrm{i}}^{(1)} C_{\mathrm{v}}^{(1)}\right) 
    \\ 
     &
     -C_{\mathrm{s}}\left(\begin{array}{c}K_{\mathrm{vs}} C_{\mathrm{v}}^{(2)} \\ K_{\mathrm{is}} C_{\mathrm{i}}^{(2)}\end{array}\right)  -C_{\mathrm{s}}\left(\begin{array}{c}
     K_{\mathrm{vs}} C_{\mathrm{v}}^{(1)} \\  K_{\mathrm{is}} C_{\mathrm{i}}^{(1)}  \end{array}\right)
\end{split}
\label{eq::Cs2}
\end{align}
\begin{align}
\begin{split}
\frac{\mathrm{d}}{\mathrm{d} t}\left(\begin{array}{c}C_{\mathrm{v}}^{(3)} \\ C_{\mathrm{i}}^{(3)}\end{array}\right)
&=-K_{\mathrm{iv}}\left(C_{\mathrm{i}}^{(0)} C_{\mathrm{v}}^{(3)}+C_{\mathrm{v}}^{(0)} C_{\mathrm{i}}^{(3)}+C_{\mathrm{i}}^{(1)} C_{\mathrm{v}}^{(2)}+C_{\mathrm{v}}^{(1)} C_{\mathrm{i}}^{(2)}\right) \\
&
 -C_{\mathrm{s}}\left(\begin{array}{c}K_{\mathrm{vs}} C_{\mathrm{v}}^{(3)} \\ K_{\mathrm{is}} C_{\mathrm{i}}^{(3)}\end{array}\right)-C_{\mathrm{s}}\left(\begin{array}{c} 
 K_{\mathrm{vs}} C_{\mathrm{v}}^{(2)} \\  K_{\mathrm{is}} C_{\mathrm{i}}^{(2)} \end{array}\right)
\end{split}
\label{eq::Cs3}
\end{align}
The corresponding equations for $C_\mathrm{s}$ are identical to those of $K_\mathrm{vs}$ and $K_\mathrm{is}$ except that the last term in each equation is the sum of those of $K_\mathrm{vs}$ and $K_\mathrm{is}$. 

\subsubsection{Perturbation on $K_{\mathrm{i v}}$}
\label{sec::Kiv}
The equations to solve for the change due to $K_{iv}$ are given in Eq~\ref{eq::Kiv1} to Eq~\ref{eq::Kiv3}. 

\begin{equation}
    \frac{\mathrm{d}}{\mathrm{d} t}\left(\begin{array}{c}C_{\mathrm{v}}^{(1)} \\ C_{\mathrm{i}}^{(1)}\end{array}\right) = -K_{\mathrm{iv}} \left (
    C_{\mathrm{i}}^{(0)} C_{\mathrm{v}}^{(1)} +
    C_{\mathrm{v}}^{(0)} C_{\mathrm{i}}^{(1)} + 
    C_{\mathrm{i}}^{(0)} C_{\mathrm{v}}^{(0)} 
    \right ) -C_{\mathrm{s}}\left(\begin{array}{c}K_{\mathrm{vs}} C_{\mathrm{v}}^{(1)} \\ K_{\mathrm{is}} C_{\mathrm{i}}^{(1)}\end{array}\right) 
\label{eq::Kiv1}
\end{equation}
\begin{align}
\begin{split}
    \frac{\mathrm{d}}{\mathrm{d} t}\left(\begin{array}{c}C_{\mathrm{v}}^{(2)} \\ C_{\mathrm{i}}^{(2)}\end{array}\right) 
    & =-K_{\mathrm{iv}}\left(
    C_{\mathrm{i}}^{(0)} C_{\mathrm{v}}^{(2)}+
    C_{\mathrm{v}}^{(0)} C_{\mathrm{i}}^{(2)}+
    C_{\mathrm{i}}^{(1)} C_{\mathrm{v}}^{(1)}+
    C_{\mathrm{i}}^{(0)} C_{\mathrm{v}}^{(1)}+
    C_{\mathrm{v}}^{(0)} C_{\mathrm{i}}^{(1)}    
    \right) \\ & 
     -C_{\mathrm{s}}\left(\begin{array}{c}K_{\mathrm{vs}} C_{\mathrm{v}}^{(2)} \\ K_{\mathrm{is}} C_{\mathrm{i}}^{(2)}\end{array}\right) 
\end{split}
\label{eq::Kiv2}
\end{align}
\begin{align}
\begin{split}
\frac{\mathrm{d}}{\mathrm{d} t}\left(\begin{array}{c}C_{\mathrm{v}}^{(3)} \\ C_{\mathrm{i}}^{(3)}\end{array}\right)
&=-K_{\mathrm{iv}}\left(
    C_{\mathrm{i}}^{(0)} C_{\mathrm{v}}^{(3)}+
    C_{\mathrm{v}}^{(0)} C_{\mathrm{i}}^{(3)}+
    C_{\mathrm{i}}^{(1)} C_{\mathrm{v}}^{(2)}+
    C_{\mathrm{v}}^{(1)} C_{\mathrm{i}}^{(2)}+
    \right. \\ & ~~~~~~~~~~~~~ \left.  
    C_{\mathrm{i}}^{(0)} C_{\mathrm{v}}^{(2)}+
    C_{\mathrm{v}}^{(0)} C_{\mathrm{i}}^{(2)}+
    C_{\mathrm{i}}^{(1)} C_{\mathrm{v}}^{(1)}
\right) \\ &
 -C_{\mathrm{s}}\left(\begin{array}{c}K_{\mathrm{vs}} C_{\mathrm{v}}^{(3)} \\ K_{\mathrm{is}} C_{\mathrm{i}}^{(3)}\end{array}\right) 
\end{split}
\label{eq::Kiv3}
\end{align}
 
\subsection{Sensitivity analysis}
The results above in Eq~\ref{eq::K01} to Eq~\ref{eq::Kiv3} can be used to predict results change on fine grids of perturbations. 
For each input parameter, we only need to solve a few more equations, then the deviation from the unperturbed solution can be calculated for as many as $\epsilon$'s. Section~\ref{sec::app} below shows results on the sensitivity of the solution on change of $K_\mathrm{vs}$ and $K_\mathrm{iv}$. It verifies that the $3^{rd}$ order perturbation captures response to $50\%$ input changes very well. It can be shown that the results in section~\ref{sec::c123} can be extended to any finite orders. Then convergence criteria can be implemented to adjust the number of correction terms automatically. 
\subsection{Uncertainty quantification}
In addition to the sensitivity analysis to each individual input parameter, 
the perturbation expansion can be applied to get aggregated uncertainty due to all the parameters. 
Denote the input parameters as $\{K_{\alpha}\}_{\alpha=1,2,3\cdots}$, 
then the deviation from unperturbed solution from each component as in Eq~\ref{eq::y+y1+...} can be summed as

\begin{align}
\begin{split}
& \left(\begin{array}{c}C_{\mathrm{v}}  \\ C_{\mathrm{i}} \end{array}\right) - 
\left(\begin{array}{c}C_{\mathrm{v}}^{(0)} \\ C_{\mathrm{i}}^{(0)}\end{array}\right) = \\ & 
\sum_{K_{\alpha}}\left[
\epsilon_{K_{\alpha}}
\left(\begin{array}{c}C_{\mathrm{v},K_{\alpha}}^{(1)} \\ C_{\mathrm{i},K_{\alpha}}^{(1)}\end{array}\right) +    
\epsilon_{K_{\alpha}}^2
\left(\begin{array}{c}C_{\mathrm{v},K_{\alpha}}^{(2)} \\ C_{\mathrm{i},K_{\alpha}}^{(2)}\end{array}\right) +    
\epsilon_{K_{\alpha}}^3
\left(\begin{array}{c}C_{\mathrm{v},K_{\alpha}}^{(3)} \\ C_{\mathrm{i},K_{\alpha}}^{(3)}\end{array}\right) + 
\cdots
\right]
\end{split}
\label{eq::yagg}
\end{align}

The perturbations $\{\epsilon_{K_{\alpha}}\}_{\alpha=1,2,3\cdots}$ can be viewed as random variables with given joint-distribution. 
Apply the variance operator on Eq~\ref{eq::yagg}, 
the aggregated uncertainty of $C_v$ and $C_i$ can be expressed as function of uncertainty of the individual input parameters and higher order correlations if any.
\section{APPLICATION} 
\label{sec::app}
We apply the above analyses to pure alpha-Fe under electron irradiation. It is reasonable to assume that only point defects are directly produced during irradiation due to the limited energy transfer between electrons and lattice atoms. In addition, we assume that no defect clustering would occur, since it necessitates a more sophisticated treatment beyond point defect kinetics. To solve Eq. \ref{eq::pk}, we use the irradiation condition and materials parameters shown in TABLE \ref{tab:params}, where $D_\mathrm{v0}$ and $D_\mathrm{i0}$ are the diffusion coefficient prefactors, and $E_{mv}$ and $E_{mi}$ are the migration barriers, respectively, and $R$ is the defect interaction distance. For simplicity, only dislocations are considered as the sinks to point defects (i.e., $C_\mathrm{s}\equiv \rho_d$, where $\rho_d$ is the dislocation density). The defect sink rate to dislocations is written as \cite{was2016fundamentals},
\begin{equation}
K_{(\mathrm{v, i}) d}=\frac{2 \pi D_{(\mathrm{v, i})}}{\ln \left(\frac{d/2}{R_{(\mathrm{v, i}) d}}\right)}, ~~\text{with dislocation distance } d=\frac{2}{\sqrt{\pi \rho_d}} 
\label{eq:kvd}
\end{equation}
The recombination rate is written as \cite{was2016fundamentals},
\begin{equation}
K_{\mathrm{i v}}=4 \pi R_{\mathrm{v, i}}\left(D_\mathrm{v}+D_\mathrm{i}\right),~~\text{where } D_{(\mathrm{v, i})}=D_{(\mathrm{v, i}) 0} \mathrm{exp}{\left(\frac{-E_{m(\mathrm{v, i})}}{k_B T}\right)}
\label{eq:kiv}
\end{equation}
\begin{table}[!ht]
\centering
\caption{Materials Parameters and radiation condition used in solving Eq. \ref{eq::pk}.}
\label{tab:params}
\begin{tabular}{|l|l|l|l|}
\hline
Displacement rate & $4.6\times10^{-3}$ dpa/s                     & Dislocation density ($\rho_d$)               & $10^{15} \mathrm{/m^2}$ \\ \hline
Lattice parameter & 0.286 nm                            & $R_{\mathrm{id}}$ (dislocation-interstitial) & 3.6 nm  \cite{short2016modeling}   \\ \hline
$D_\mathrm{v0}$   & $8.016\times10^{-7} \mathrm{m^2/s}$ \cite{short2016modeling} & $R_{\mathrm{vd}}$ (dislocation-vacancy)      & 1.2 nm  \cite{short2016modeling}  \\ \hline
$D_\mathrm{i0}$   & $2.09\times 10^{-7}\mathrm{ m^2/s}$ \cite{short2016modeling}& $R_{\mathrm{iv}}$ (intersitial-vacancy)     & 0.65 nm \cite{meslin2008cluster}   \\ \hline
$E_{mv}$          & 0.86  eV \cite{short2016modeling}                          & Temperature ($T$)                           & 300 K                   \\ \hline
$E_{mi}$          & 0.17 eV \cite{short2016modeling}                            & \textbf{}                                   &                         \\ \hline
\end{tabular}
\end{table}
\begin{figure}[h]
\centering
\includegraphics[width=0.55\textwidth]{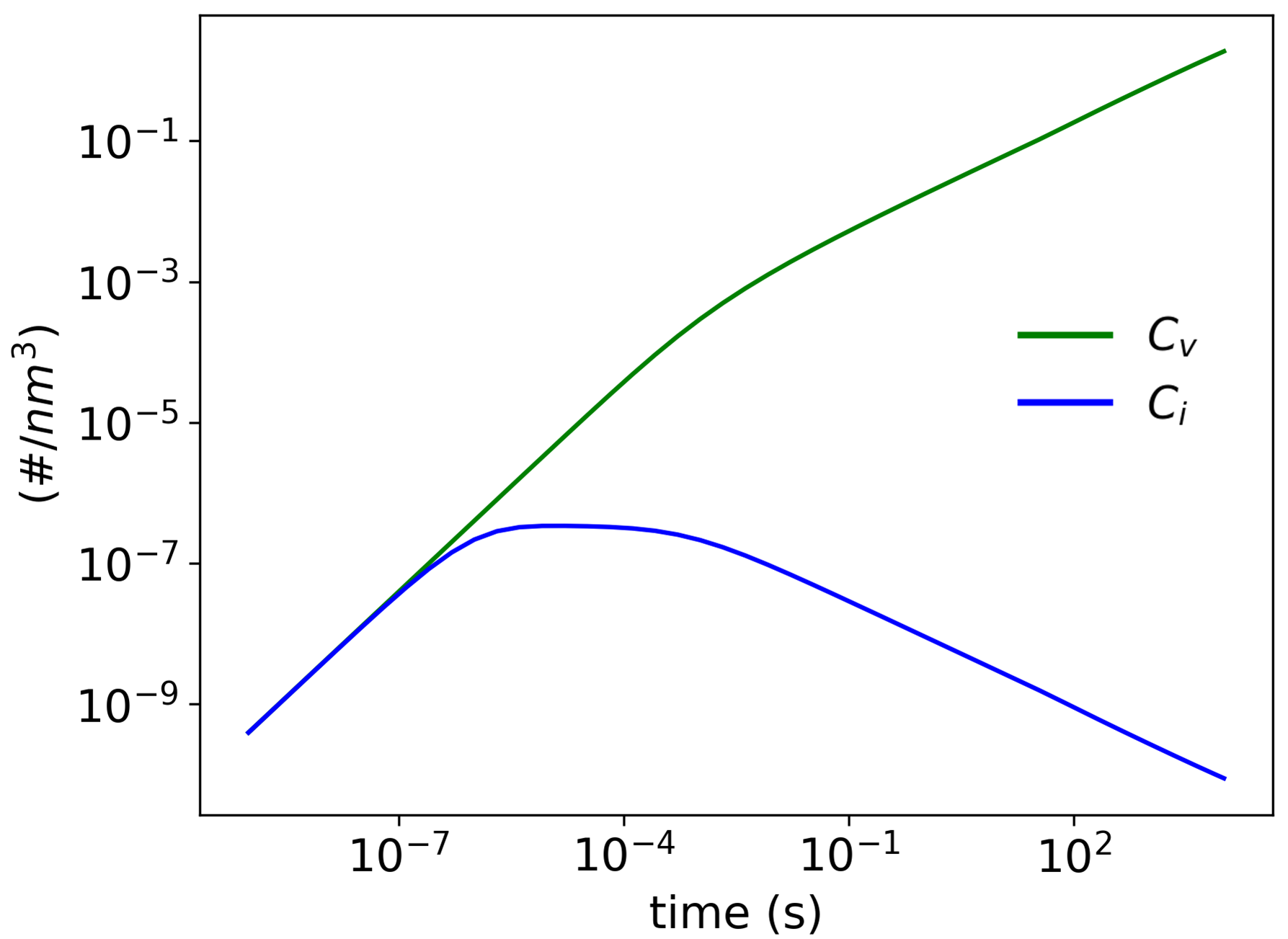}
\caption{$C_\mathrm{v}$ and $C_\mathrm{i}$ vs. time given the settings in Table \ref{tab:params}.}
\label{fig:Cv_Ci}
\end{figure}

The solution of Eq. \ref{eq::pk} is shown in Figure \ref{fig:Cv_Ci}, which falls into the regime of low-temperature high-sink density scenario \cite{was2016fundamentals} and the steady state has not yet reached until 10,000 s. To validate our analyses, we consider the variations in $K_{\mathrm{iv}}$ and $K_{\mathrm{vs}}$, which depend on fundamental defect properties. Uncertainty can be introduced by various factors, such as impurities, local stress, and computation accuracy \cite{hashimoto2014effect,fu2005ab}. Here, the uncertainty range is considered to be up to 50\%. This choice is based on the observation of Short et al.'s work \cite{short2016modeling}, where it was shown that a slight change (0.01 eV) in vacancy migration energy can cause a significant change in the vacancy concentration profile. Here, we translate this variation into the rate constants, by estimating the change in the vacancy diffusion coefficient given the Arrhenius form. Given the formulations provided in Eqs. \ref{eq:kvd} and \ref{eq:kiv}, it leads to $\mathrm{exp}(0.01\mathrm{eV}/0.025851\mathrm{eV})=1.47$ or 47\% change in $K_\mathbf{vs}$ and $K_\mathbf{iv}$ at 300 K. As another example, typical DFT convergence inaccuracy around 5 meV would translate to $\mathrm{exp}(0.005\mathrm{eV}/0.0253\mathrm{eV})=1.21$ or 21\% change in $K_\mathbf{vs}$ and $K_\mathbf{iv}$ at 300 K. Hence, in the following demonstrations, we show the results for variations of $K_\mathbf{vs}$ and $K_\mathbf{iv}$ at $\pm20\%$ and $\pm50\%$ changes. To simplify notation, $\alpha\equiv\Delta K_\mathbf{iv}/K_\mathbf{iv}$ and $\beta\equiv\Delta K_\mathbf{vs}/K_\mathbf{vs}$.

\begin{figure}[h]
\centering
\includegraphics[width=0.95\textwidth]{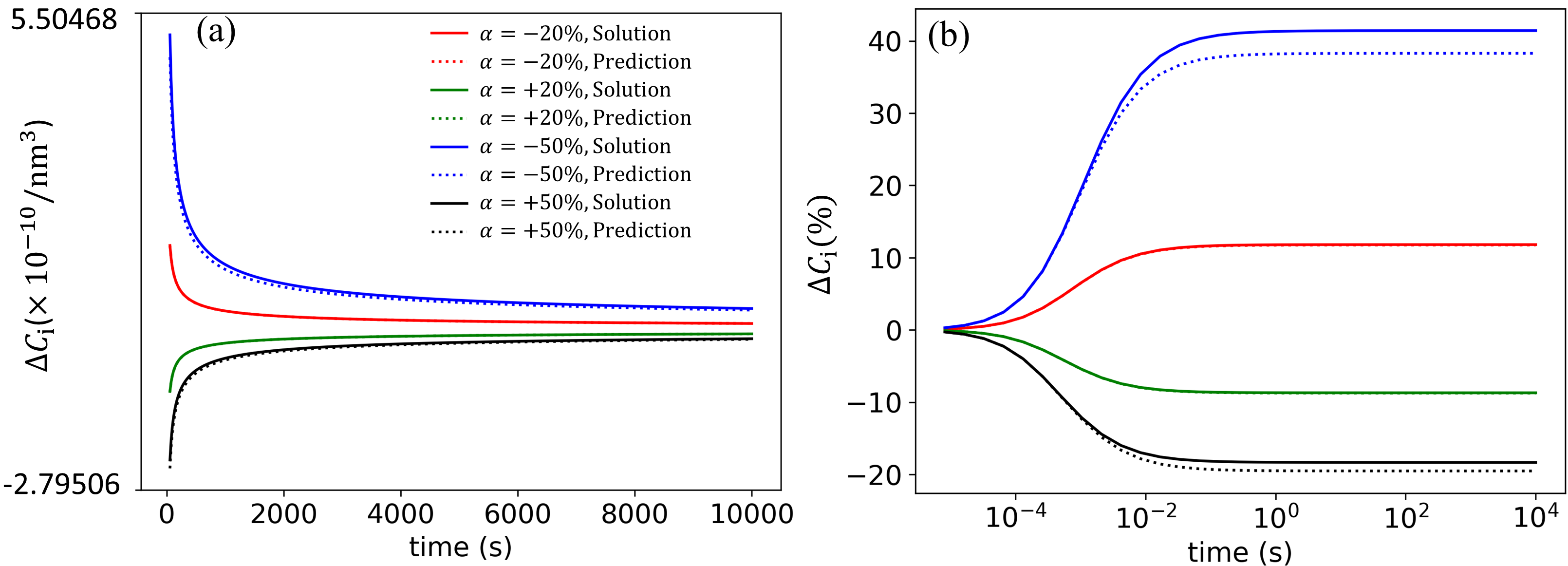}
\caption{(a) $\Delta C_\mathrm{i}$ vs. time under $\alpha=\pm 20\%$ and  $\alpha=\pm 50\%$. Solid lines represent the solution from solving Eq. \ref{eq::pk}. Dotted lines represent the results up to third-order correction based on Eqs~\ref{eq::Kiv1},\ref{eq::Kiv2},\ref{eq::Kiv3}. (b) shows the corresponding percentage changes.}
\label{fig:Ci_Kiv}
\end{figure}

\begin{figure}[h]
\centering
\includegraphics[width=0.95\textwidth]{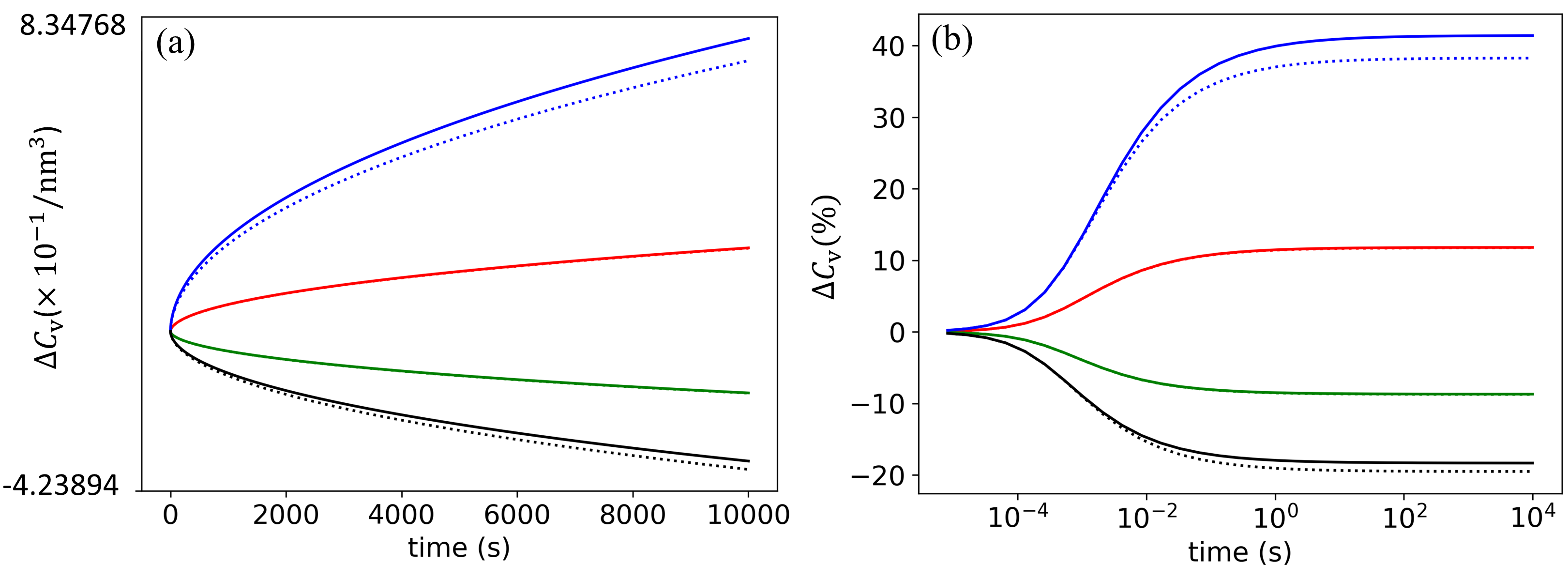}
\caption{
(a) $\Delta C_\mathrm{v}$ vs. time under $\alpha=\pm 20\%$ and  $\alpha=\pm 50\%$. Solid lines represent the solution from solving Eq. \ref{eq::pk}. Dotted lines represent the results up to third-order correction based on Eqs~\ref{eq::Kiv1},\ref{eq::Kiv2},\ref{eq::Kiv3}. (b) shows the corresponding percentage changes.
The color coding is the same as Fig~\ref{fig:Ci_Kiv}. 
}
\label{fig:Cv_Kiv}
\end{figure}

\begin{figure}[h]
\centering
\includegraphics[width=0.55\textwidth]{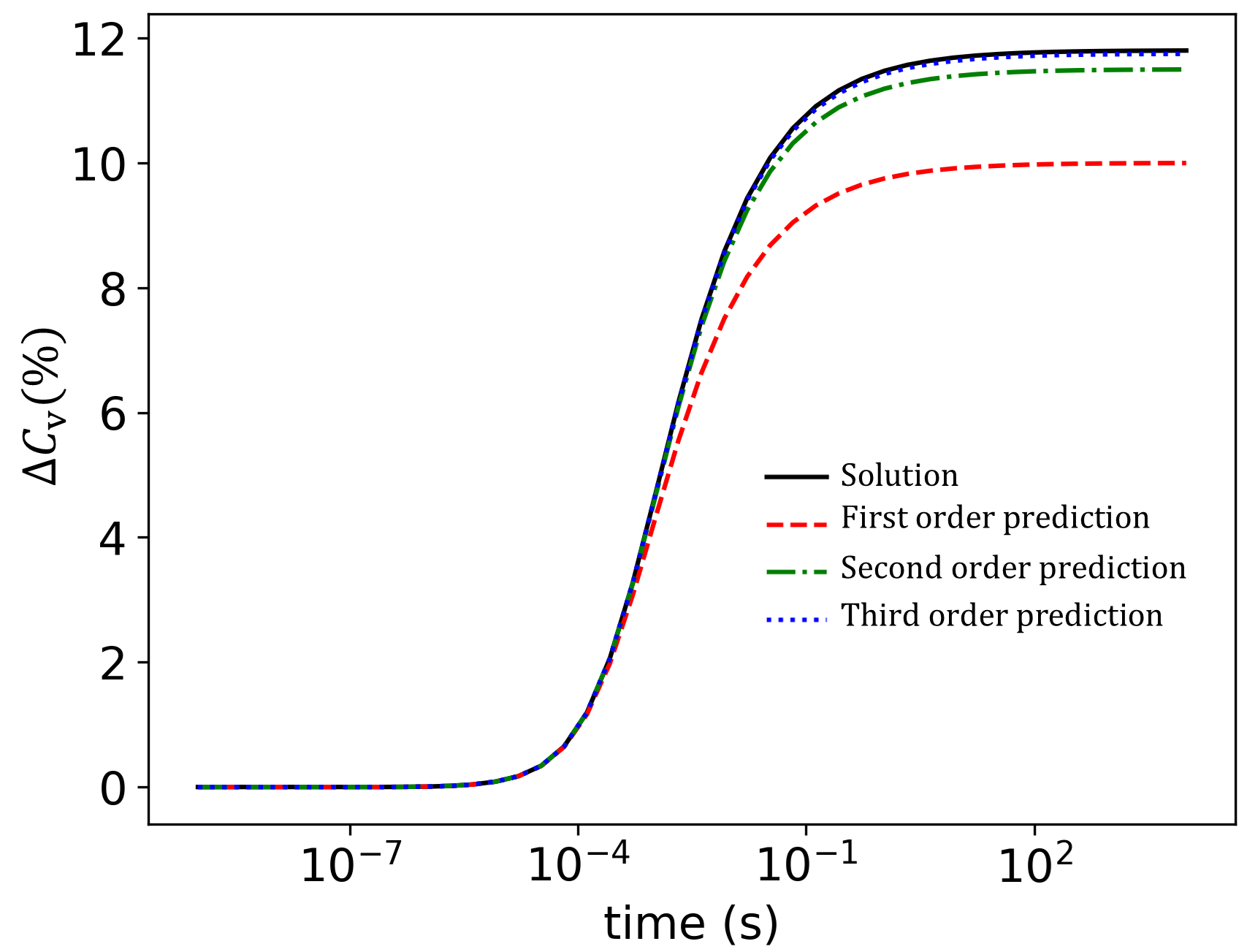}
\caption{
$\Delta C_\mathrm{v}(\%)$ vs. time under $\alpha=-20\%$. The solid line represents the solution from solving Eq~\ref{eq::pk}. 
The dashed line represents the results up to first-order correction based on Eq~\ref{eq::Kiv1}. 
The dashdot line represents the results up to second-order correction based on Eq~\ref{eq::Kiv2}.
The dotted line represents the results up to third-order correction based on Eq~\ref{eq::Kiv3}. 
}
\label{fig:Cv_Kiv_order}
\end{figure}

First, we consider the independent uncertainty in $K_\mathbf{iv}$ and $K_\mathbf{vs}$. Figures \ref{fig:Ci_Kiv} and \ref{fig:Cv_Kiv} plot the real and percentage changes in $C_\mathrm{v}$ and $C_\mathrm{i}$ via direct solving Eq. \ref{eq::pk} with changed $K_\mathbf{iv}$ and applying the perturbation analysis up to third-order correction. Since $K_\mathbf{iv}$ indicates the loss due to recombination, a more negative $\alpha$ would lead to higher defect concentration ($\Delta C>0$), and vice versa (Figures \ref{fig:Ci_Kiv}a and \ref{fig:Cv_Kiv}a). Note that the positive and negative $\alpha$ don't exhibit  symmetry in $\Delta C_\mathbf{i}$ vs. time due to the nonlinear nature of the two coupled equations. It can also been seen that the absolute discrepancy in $C_\mathrm{i}$ and $C_\mathrm{v}$ embraces an opposite trend, where the former decreases (Figure \ref{fig:Ci_Kiv}a) while the latter increases (Figure \ref{fig:Cv_Kiv}a) with time. However, the relative discrepancy increases for both interstitial and vacancy concentrations (Figures \ref{fig:Ci_Kiv}b and \ref{fig:Cv_Kiv}b). In all cases considered, the third-order correction agrees well with the direct solution, up to 50 \% variation in the $K_\mathbf{iv}$. With even higher variations, higher order corrections are needed as third-order exhibits certain difference with the direct solution. To reveal how the order of correction affect the accuracy in prediction, Figure \ref{fig:Cv_Kiv_order} displays $\Delta C_\mathbf{v}$ at the first, second and third order corrections, under $\alpha=-20\%$. A large deviation exists for the first order correction, however, the second order can already capture the solution very well and the third order agrees perfectly with the solution.

\begin{figure}[h!]
\centering
\includegraphics[width=0.95\textwidth]{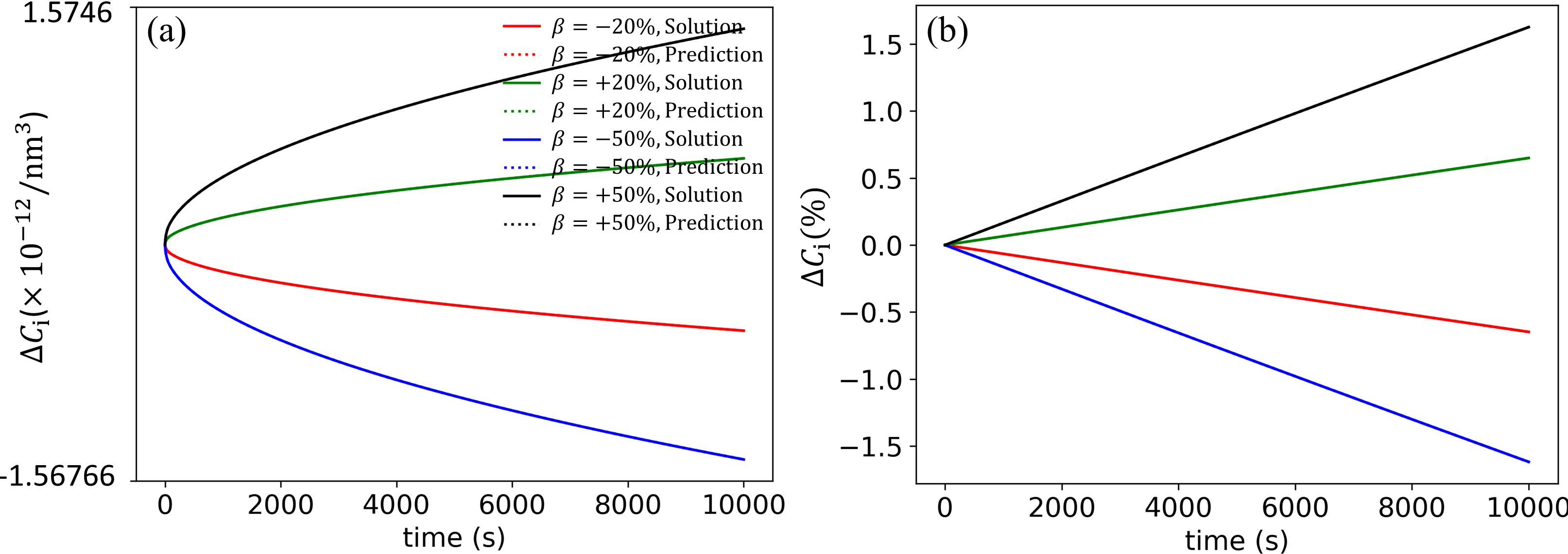}
\caption{
(a) $\Delta C_\mathrm{i}$ vs. time under $\beta=\pm 20\%$ and  $\beta=\pm 50\%$. Solid lines represent the solution from solving Eq. \ref{eq::pk}. Dotted lines represent the results up to third-order correction based on Eqs~\ref{eq::Kvs1},\ref{eq::Kvs2},\ref{eq::Kvs3}. (b) shows the corresponding percentage changes.
}
\label{fig:Ci_Kvs}
\end{figure}

\begin{figure}[h]
\centering
\includegraphics[width=0.95\textwidth]{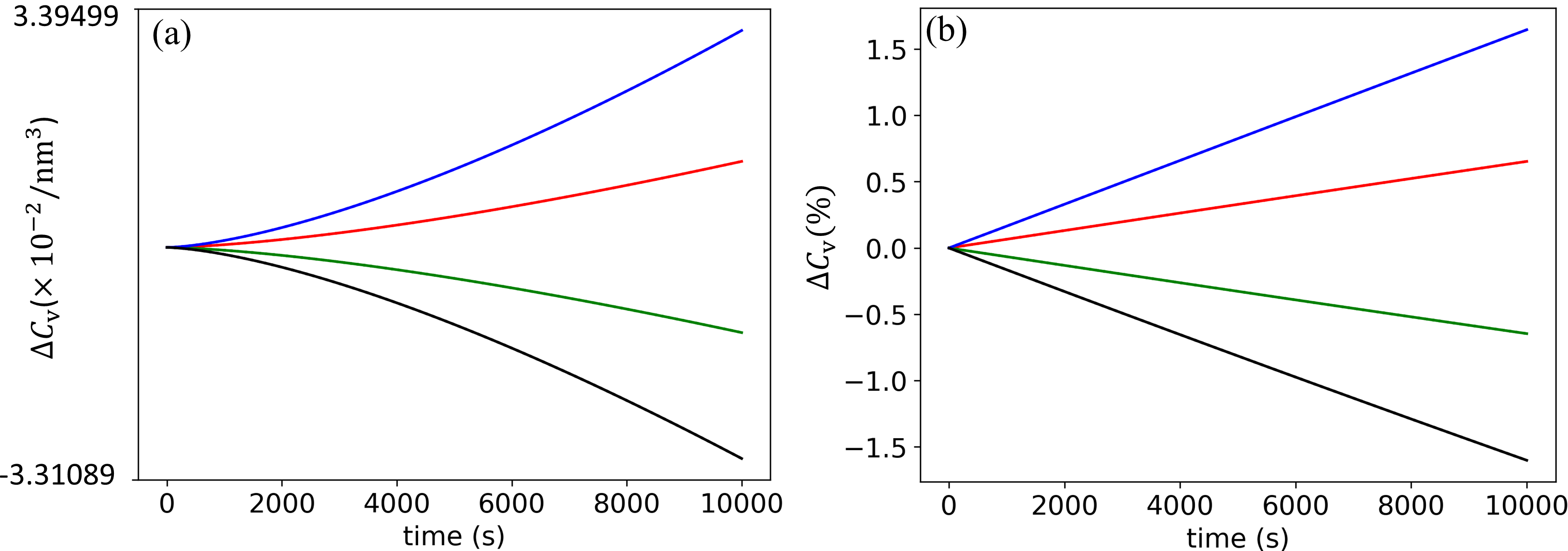}
\caption{
(a) $\Delta C_\mathrm{v}$ vs. time under $\beta=\pm 20\%$ and  $\beta=\pm 50\%$. Solid lines represent the solution from solving Eq. \ref{eq::pk}. Dotted lines represent the results up to third-order correction based on Eqs~\ref{eq::Kvs1},\ref{eq::Kvs2},\ref{eq::Kvs3}. (b) shows the corresponding percentage changes.
The color coding is the same as Fig~\ref{fig:Ci_Kvs}. 
}
\label{fig:Cv_Kvs}
\end{figure}

\begin{figure}[ht!]
\centering
\includegraphics[width=0.55\textwidth]{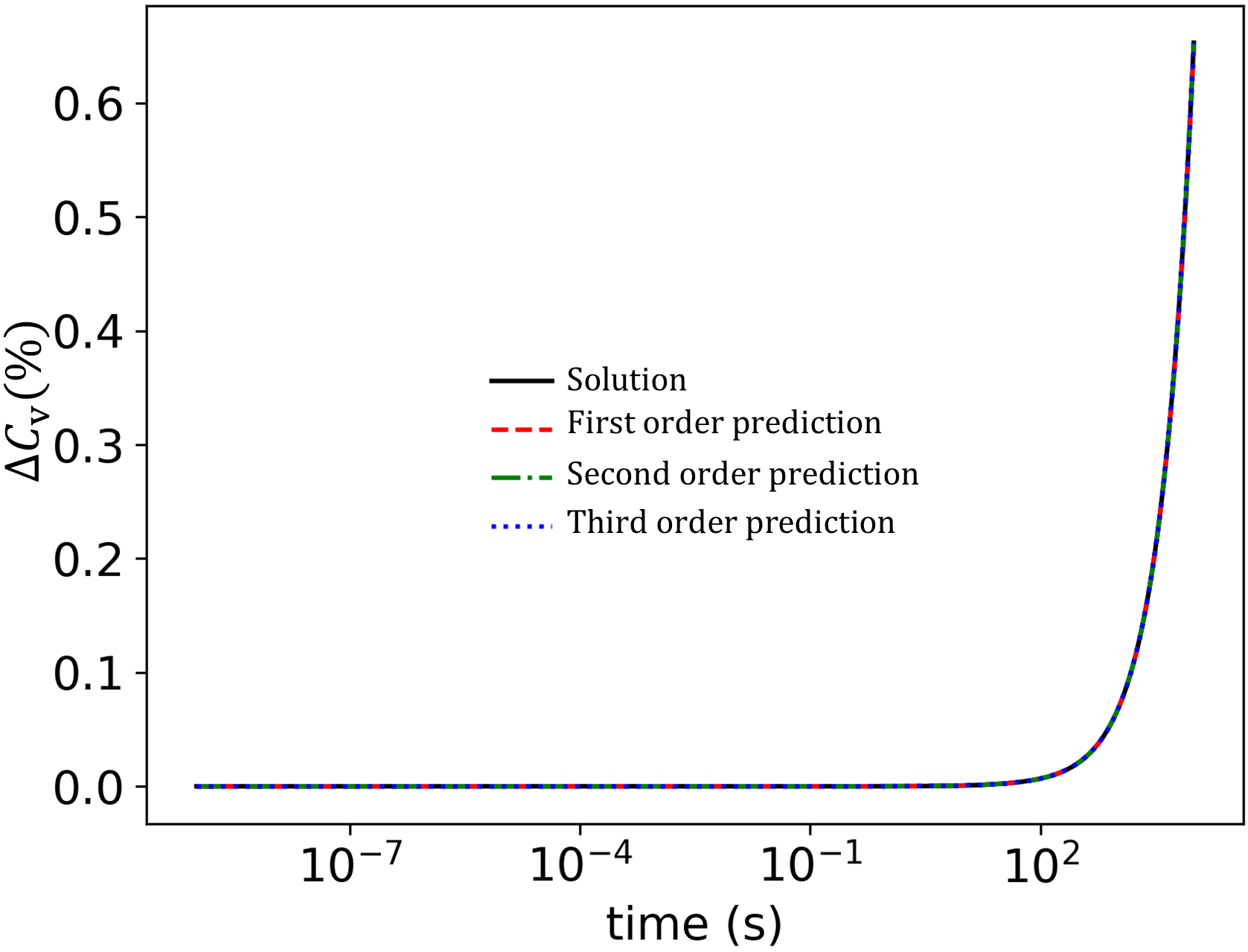}
\caption{
$\Delta C_\mathrm{v}(\%)$ vs. time under $\beta=-20\%$. The solid line represents the solution from solving Eq~\ref{eq::pk}. 
The dashed line represents the results up to first-order correction based on Eq~\ref{eq::Kvs1}. 
The dashdot line represents the results up to second-order correction based on Eq~\ref{eq::Kvs2}.
The dotted line represents the results up to third-order correction based on Eq~\ref{eq::Kvs3}. 
}
\label{fig:Cv_Kvs_order}
\end{figure}

The effect of varying $K_\mathbf{vs}$ on $C_\mathrm{i}$ and $C_\mathrm{v}$ is shown in Figures \ref{fig:Ci_Kvs} and \ref{fig:Cv_Kvs}, which show a lessened impact compared to that in $K_\mathbf{iv}$. Both absolute and relative discrepancy in defect concentrations due to uncertainty in $K_\mathbf{vs}$, increases with time. Note that, $\Delta C_\mathrm{i}$ and $\Delta C_\mathrm{v}$ exhibit an opposite sign for a given $\beta$. For example, with 50 \% increased $K_\mathbf{vs}$ ($\beta = 50\%$), $\Delta C_\mathrm{v}$ become increasingly negative with time. Due to the reduction in $C_\mathrm{v}$, less recombination causes an increase in $C_\mathrm{i}$, hence, $\Delta C_\mathrm{i}$ become increasingly positive with time. The relative changes demonstrate the same trend to the real changes. In these cases, third order corrections can fully predict the direct solution. From Figure \ref{fig:Cv_Kvs_order} with $\beta=-20\%$ ($\beta=\pm50\%$ were also evaluated, exhibiting the same behavior), it suggests that even first order correction is capable to capture all discrepancies.

\begin{figure}[]
\centering
\includegraphics[width=0.95\textwidth]{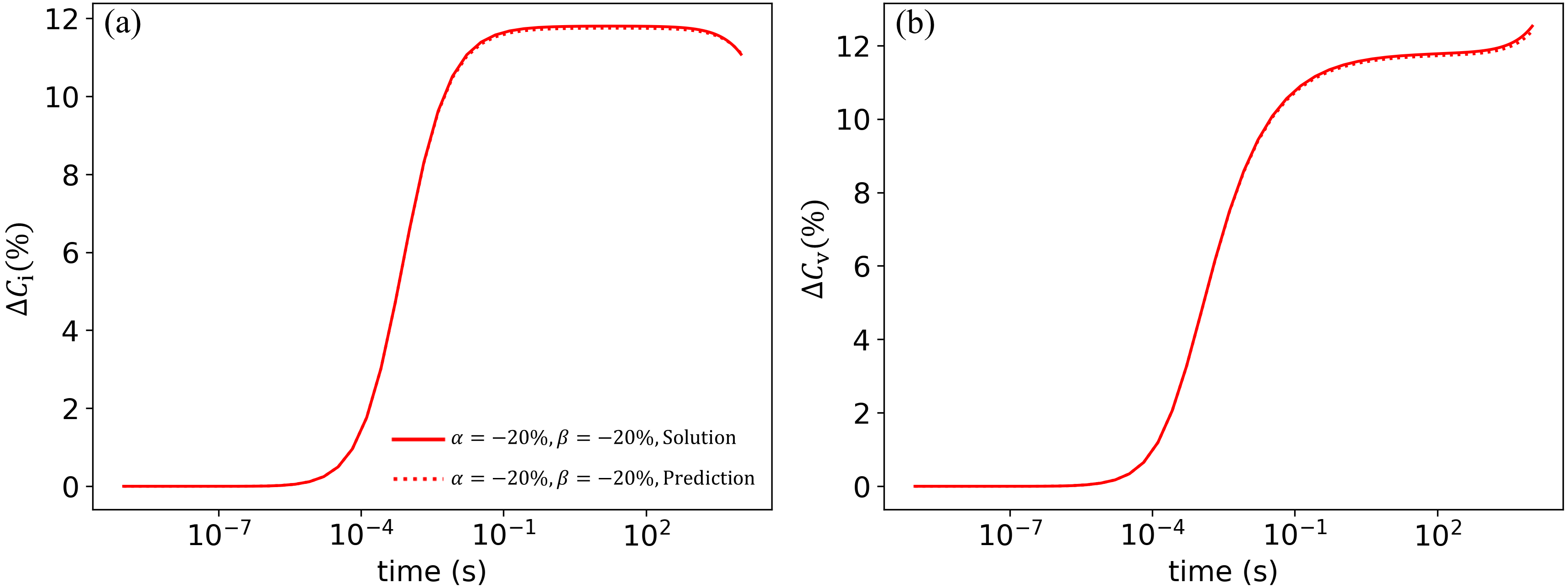}
\caption{
$\Delta C_\mathrm{i}(\%)$ (a) and $\Delta C_\mathrm{v}(\%)$ (b) vs. time under changes of both $K_{iv}$ and $K_{vs}$ ($\alpha=-20\%$ and  $\beta=-20\%$). Solid lines represent the solution from solving Eq. \ref{eq::pk} with new $K_{iv}$ and $K_{vs}$ values. Dotted lines represent the prediction from adding the results of $K_{iv}$ induced change (Eqs~\ref{eq::Kiv1},\ref{eq::Kiv2},\ref{eq::Kiv3}) and $K_{vs}$ induced change (Eqs~\ref{eq::Kvs1},\ref{eq::Kvs2},\ref{eq::Kvs3}). 
}
\label{fig:Cv_Ci_Kiv_Kvs}
\end{figure}

Finally, we evaluate the simultaneous variation in both $K_\mathbf{iv}$ and $K_\mathbf{vs}$ given the same dependence on the vacancy diffusion coefficient. Figure \ref{fig:Cv_Ci_Kiv_Kvs} shows the relative changes in $C_\mathrm{i}$ and $C_\mathrm{v}$ under $\alpha=-20\%$ and $\beta=-20\%$. Third-order prediction overlaps the direct solution at all times, indicating the strong efficacy of this perturbation methodology for uncertainty in multiple parameters. Note, the other two rates ($K_0$ and $K_\mathbf{is}$) can be similarly considered for relevant scenarios involving uncertainty in dose rate and the interstitial-dislocation interactions. 

\section{CONCLUSION}
We derived the perturbation expansion to analyze the response of the point defect kinetics equation to the uncertainty and sensitivity in the input parameters. 
The results were numerically verified on parameters $K_\mathrm{vs}$ and $K_\mathrm{iv}$ up to 50 \% variations, considering the case of electron irradiated pure $\alpha$-Fe. The method has the advantage that by solving a few extra equations, the sensitivity analysis can be performed on continuously changing parameters, instead of solving the original equation repeatedly on all the cases. We also discussed the capability of the analyses to generate aggregated uncertainty due to uncertainty in multiple rate constants. This method can be extended to add higher orders adaptively if substantial uncertainty exists in those rate constants. 
 
\textbf{ACKNOWLEDGEMENTS}: We acknowledge the support from the Department of Nuclear Engineering at Penn State University.

\setlength{\baselineskip}{11pt}
\bibliographystyle{mc2023}
\bibliography{mc2023}

\end{document}